\UseRawInputEncoding
\documentclass[12pt,prd,onecolumn,floatfix,nofootinbib]{revtex4}
\usepackage{amssymb}
\usepackage{amsmath}
\usepackage{graphicx,color,dcolumn,booktabs,bm}
\usepackage{longtable,lscape}
\usepackage{txfonts}
\usepackage{overpic}
\usepackage{indentfirst}
\usepackage{cases}
\usepackage{multirow}
\usepackage{epstopdf}
\usepackage{ulem}
\usepackage[colorlinks,
            citecolor=blue,
            anchorcolor=red,
            menucolor=red,
            linkcolor=red,
            filecolor=red,
            runcolor=red,
            urlcolor=blue,
            frenchlinks=false]{hyperref}

\allowdisplaybreaks
\begin{document}

\title{Doubly heavy tetraquarks: heavy quark bindings and chromomagnetically
mixings}
\author{Xin-Yue Liu$^{1}$}
\email{Liuxinyue.sun@qq.com}
\author{Wen-Xuan Zhang$^{1}$}
\email{zhangwx89@outlook.com}
\author{Duojie Jia$^{1,2}$ \thanks{}}
\email{jiadj@nwnu.edu.cn; corresponding author}
\affiliation{$^1$Institute of Theoretical Physics, College of Physics and Electronic
Engineering, Northwest Normal University, Lanzhou 730070, China \\
$^2$Lanzhou Center for Theoretical Physics, Lanzhou University, Lanzhou,
730000,China \\
}

\begin{abstract}
We introduce an enhanced binding energy $B_{QQ}$ between heavy-heavy quarks $%
QQ$ and a flux-tube correction into the chromomagnetic interaction model to
study nonstrange doubly-heavy tetraquarks $T_{QQ}$ ($Q=c,b)$. A simple
relation in terms of baryon masses is proposed to estimate the binding
energies $B_{QQ}$ and thereby map the flux-tube corrections in doubly-heavy
tetraquarks $T_{cc}$, $T_{bb}$ and $T_{bc}$. Our computation via
diagonalization of chromomagnetic interaction predicts the doubly charmed
tetraquark $T_{cc}$ (in color rep. $\bar{3}_{c}\otimes 3_{c}$) and $%
T_{cc}^{\ast }$ (in $6_{c}\otimes \bar{6}_{c}$) with $IJ^{P}=01^{+}$ to have
masses of $3879.2$ MeV and $4287.6$ MeV, respectively, with the former
being in consistent with the measured mass $3874.7\pm 0.05$ MeV of the
doubly charmed tetraquark $T_{cc}(1^{+})=cc\bar{u}\bar{d}$ discovered by
LHCb. Further mass predictions are given of the doubly bottom tetraquarks $%
T_{bb}$ and the bottom-charmed tetraquarks $T_{bc}$ with $%
J^{P}=0^{+},1^{+},2^{+}$ and $I=0,1$. A chromomagnetical mixing between the
color configurations $\bar{3}_{c}\otimes 3_{c}$ and $6_{c}\otimes \bar{6}_{c}
$ is noted for the bottom-charmed states $T_{bc}$ with $IJ^{P}=01^{+}$ and $%
IJ^{P}=1(0^{+},1^{+})$.
\end{abstract}

\maketitle
\date{\today }

\section{Introduction}

Multiquarks like tetraquarks (with quark configuration $q^{2}\bar{q}^{2}$)
and pentaquarks ($q^{4}\bar{q}$) had been suggested earlier at the birth of
the quark model \cite{Gellman64,Zweig64} long before the advent of quantum
chromodynamics (QCD). The multiquarks are considered to be exotic as they do
not fit into conventional scheme of hadron classification via meson and
baryons whereas they are allowed by QCD in principle. The first explicit
calculation of multiquark states was carried out in the 1970s based on the
MIT bag model \cite{Jaffe77a,Jaffe77b}. These and other early theoretical
exporations triggered a lot of experimental searches with no conlusive
results until 2003 when the Belle discovered the first exotic hadron $%
X(3872) $ \cite{BelleX3872}.

Since the discovery of the $X(3872)$ which was confirmed by other
experiments \cite{Workman:2022ynf}, many tetraquark candidates have been
observed later, which include the charmonium-like states the $Z_{c}(3900)$
\cite{BESIII collaboration23} and many others \cite{Workman:2022ynf}.
Recently, the LHCb collaboration reported a discovery of the the first
tetraquak with two charm quarks, the $X(3875)$ state ($IJ^{P}=01^{+}$) \cite%
{LHCb:2021vvq,LHCb:2021auc}. This observation arouses interest of the doubly
heavy(DH) tetraquarks $T_{QQ}$($Q=c,b$) and led to a set of studies on them
based on pictures of hadronic molecular \cite{G.-J. Ding471,R. Molina405,H.
Xu472,L. Tang473}, and of the compact tetraquark \cite{J. L. Ballot446,S.
Zouzou447,J. Carlson448,X. Yan450,Q. Meng2022454,C. Deng455}, \cite{S. S.
Agaev459,G. Q. Feng460,J.-B. Cheng463}. See Refs. \cite%
{Ali:2017,KRS:2018,LCCL:2019} for reviews.

The purpose of this work is to introduce a binding energy $B_{QQ}$ between
heavy-heavy quarks to study DH tetraquarks $T_{QQ}$ in the framework of
chromomagnetic interaction model. We use a simple relaion connecting the
measured masses of baryons to estimate $B_{QQ}$ and compute the ground-state
masses of the nonstrange DH tetraquarks $T_{cc}$, $T_{bb}$ and $T_{bc}$ via
diagonalization of the chromomagnetic interaction (CMI). For the masses of
the doubly charmed tetraquark we predict $M(T_{cc},1^{+})=3879.2$ MeV and $%
M(T_{cc}^{\ast },1^{+})=4287.6$ MeV, with the former being in consistent
with that of the LHCb-measured mass $3874.7\pm 0.05$ MeV. The further
compution of masses of the DH tetraquarks $T_{bb}$ and $T_{bc}$ is performed
similarly. A discussion is given for the chromomagnetical mixing among the
color-spin states of the tetraquarks with color configurations $\bar{3}%
_{c}\otimes 3_{c}$ and $6_{c}\otimes \bar{6}_{c}$, mainly occured for the
bottom-charmed states $T_{bc}$ with $IJ^{P}=01^{+}$ and $%
IJ^{P}=1(0^{+},1^{+})$.

After introduction, in Sect.~\ref{model}, we describe the CMI model for the
DH tetraquarks and classify wavefunctions of the nonstrange DH tetraquarks
via their symmetry in color and spin-falvor spaces. In Sect.~\ref{Parameters}%
, three sets of the parameters involved in the CMI model are determined,
including the binding energies. The numerical results are given in details
for all nonstrange DH tetraquarks $T_{cc}$, $T_{bb}$ and $T_{bc}$ in Sec.~%
\ref{results}. We end with conclusions and remarks in Sec.~\ref{Conclusions}.

\section{Chromomagnetic Interaction Model}

\label{model}

For ground state of a multiquark $T_{QQ}=QQ\bar{q}\bar{q}$($Q=c,b,q=u,d$),
the chromomagnetic interaction model is given by the Sakharov--Zeldovich
formula \cite{SakharovZ:66,Zur:AP08}
\begin{equation}
M=\sum_{i}\left( m_{i}+E_{i}\right) +B_{QQ}+\langle H_{CMI}\rangle ,
\label{Mass}
\end{equation}%
where $m_{i}$ ($i=1,\cdots ,4$) is the effective mass of the $i$-th quark in
hadron which can be the charm($c$), bottom($b$) or the light nonstrange($%
q=u,d$) quark, $E_{i}$ is the effective energy of color flux attached to the
$i$-th quark, $B_{QQ}$ stands for enhanced binding energy between two heavy
quarks $QQ$($=cc,bb,bc$) which enters here to account for an extra binding
between $QQ$ compared to that among the light quarks \cite{KR:D14}, and $%
H_{CMI}$ is the chromomagnetic interaction in hadron, given by the
Hamiltonian \cite{RujulaGG:75}
\begin{equation}
H_{CMI}=-\sum_{i<j}\left( \mathbf{\lambda }_{i}\cdot \mathbf{\lambda }%
_{j}\right) \left( \mathbf{\sigma }_{i}\cdot \mathbf{\sigma }_{j}\right)
\frac{A_{ij}}{m_{i}m_{j}}.  \label{HCMI}
\end{equation}%
Here, $\mathbf{\lambda }_{i}$ is the Gell-Mann matrices and $\mathbf{\sigma }%
_{i}$ stand for the Pauli matrices associated with the $i$-th quark, or the $%
i$-th anti-quark for which $\mathbf{\lambda }_{i}$ should be replaced by $-%
\mathbf{\lambda }_{i}$. The averaged matrix $\langle H_{CMI}\rangle $ of the
CMI in Eq. (\ref{Mass}), evaluated in tetraquark configurations, has been
simplified to the hamiltonian Eq. (\ref{HCMI}) in color-spin space, with the
pararameter $A_{ij}$ describes the effective CMI coupling between the $i$-th
(anti-)quark and the $j$-th (anti-)quark.

In order to find the spin-multiplets and their state mixings due to the CMI
in Eq. (\ref{Mass}), one has to exhaust all possible spin and color
wavefunctions of a tetraquark system and combine them appropriately with the
flavor configurations, so that they satisfy the constraint of the
flavor-color--spin symmetry from Pauli principle. Obtaining the
wavefunctions, one can calculate the matrix elements of $H_{CMI}$ using the
approach illustrated in Refs. \cite{LuoCLL:2017,Zhang:2021yul}.

For the spin configuration of the tetraquark $QQ\bar{q}\bar{q}$, the
possible wavefunctions (denoted by $\chi ^{T}$) are
\begin{equation}
\begin{aligned} \chi_{1}^{T}={\left| {\left(Q_{1}Q_{2}\right)}_{1}
{\left(\bar{q}_{3}\bar{q}_{4}\right)}_{1} \right\rangle}_{2}, \quad
\chi_{2}^{T}={\left| {\left(Q_{1}Q_{2}\right)}_{1}
{\left(\bar{q}_{3}\bar{q}_{4}\right)}_{1} \right\rangle}_{1}, \\
\chi_{3}^{T}={\left| {\left(Q_{1}Q_{2}\right)}_{1}
{\left(\bar{q}_{3}\bar{q}_{4}\right)}_{1} \right\rangle}_{0}, \quad
\chi_{4}^{T}={\left| {\left(Q_{1}Q_{2}\right)}_{1}
{\left(\bar{q}_{3}\bar{q}_{4}\right)}_{0} \right\rangle}_{1}, \\
\chi_{5}^{T}={\left| {\left(Q_{1}Q_{2}\right)}_{0}
{\left(\bar{q}_{3}\bar{q}_{4}\right)}_{1} \right\rangle}_{1}, \quad
\chi_{6}^{T}={\left| {\left(Q_{1}Q_{2}\right)}_{0}
{\left(\bar{q}_{3}\bar{q}_{4}\right)}_{0} \right\rangle}_{0}, \end{aligned}
\label{spinT}
\end{equation}%
where the subscripts of numbers denote spins of the heavy diquark, the light
antidiquark, and the tetraquark state.

Based on the color $SU(3)_{c}$ symmetry, one can obtain two combinations of
color singlets $6_{c}\otimes \bar{6}_{c}$ and $\bar{3}_{c}\otimes 3_{c}$ for
tetraquarks, which are

\begin{eqnarray}
\phi _{1}^{T} &=&{\left\vert {\left( Q_{1}Q_{2}\right) }^{6}{\left( \bar{q}%
_{3}\bar{q}_{4}\right) }^{\bar{6}}\right\rangle =}\frac{1}{\sqrt{6}}\left( rr%
\bar{r}\bar{r}+gg\bar{g}\bar{g}+bb\bar{b}\bar{b}\right)  \notag \\
&&+\frac{1}{2\sqrt{6}}\left( rb\bar{b}\bar{r}+br\bar{b}\bar{r}+gr\bar{g}\bar{%
r}+rg\bar{g}\bar{r}+gb\bar{b}\bar{g}+bg\bar{b}\bar{g}\right.  \notag \\
&&+gr\bar{r}\bar{g}+rg\bar{r}\bar{g}+gb\bar{g}\bar{b}\left. +bg\bar{g}\bar{b}%
+rb\bar{r}\bar{b}+br\bar{r}\bar{b}\right) ,  \label{ph1}
\end{eqnarray}%
\begin{eqnarray}
\phi _{2}^{T} &=&{\left\vert {\left( Q_{1}Q_{2}\right) }^{\bar{3}}{\left(
\bar{q}_{3}\bar{q}_{4}\right) }^{3}\right\rangle }  \notag \\
&=&\frac{1}{2\sqrt{3}}\left( rb\bar{b}\bar{r}-br\bar{b}\bar{r}-gr\bar{g}\bar{%
r}+rg\bar{g}\bar{r}+gb\bar{b}\bar{g}-bg\bar{b}\bar{g}\right.  \notag \\
&&\left. +gr\bar{r}\bar{g}-rg\bar{r}\bar{g}-gb\bar{g}\bar{b}+bg\bar{g}\bar{b}%
-rb\bar{r}\bar{b}+br\bar{r}\bar{b}\right) .  \label{ph2}
\end{eqnarray}

For the nonstrange tetraquarks with given heavy pair $QQ$, the isovector
states $QQ\{\bar{n}\bar{n}\}_{f}$ and the isoscalar states $QQ[\bar{n}\bar{n}%
]_{f}$ do not mix since we ingore isospin breaking effects. So, we are
position to combine the flavor, color and spin wavefunctions together
provided that the constraint from the Pauli principle is imposed. In the
diquark--antidiquark picture, there are twelve possible bases for the
wavefunction, which are in terms of the notation $|(QQ)_{spin}^{Color}(\bar{q%
}\bar{q})_{spin}^{Color}\rangle $ in diquark-antidiquark picture,

\begin{equation}
\begin{aligned} \phi_{1}^{T}\chi_{1}^{T}={\left|
{\left(Q_{1}Q_{2}\right)}_{1}^{6}
{\left(\bar{q}_{3}\bar{q}_{4}\right)}_{1}^{\bar{6}} \right\rangle}_{2}
\delta_{12}^{A}\delta_{34}^{A}, \\ \phi_{2}^{T}\chi_{1}^{T}={\left|
{\left(Q_{1}Q_{2}\right)}_{1}^{\bar{3}}
{\left(\bar{q}_{3}\bar{q}_{4}\right)}_{1}^{3} \right\rangle}_{2}
\delta_{12}^{S}\delta_{34}^{S}, \\ \phi_{1}^{T}\chi_{2}^{T}={\left|
{\left(Q_{1}Q_{2}\right)}_{1}^{6}
{\left(\bar{q}_{3}\bar{q}_{4}\right)}_{1}^{\bar{6}} \right\rangle}_{1}
\delta_{12}^{A}\delta_{34}^{A}, \\ \phi_{2}^{T}\chi_{2}^{T}={\left|
{\left(Q_{1}Q_{2}\right)}_{1}^{\bar{3}}
{\left(\bar{q}_{3}\bar{q}_{4}\right)}_{1}^{3} \right\rangle}_{1}
\delta_{12}^{S}\delta_{34}^{S}, \\ \phi_{1}^{T}\chi_{3}^{T}={\left|
{\left(Q_{1}Q_{2}\right)}_{1}^{6}
{\left(\bar{q}_{3}\bar{q}_{4}\right)}_{1}^{\bar{6}} \right\rangle}_{0}
\delta_{12}^{A}\delta_{34}^{A}, \\ \phi_{2}^{T}\chi_{3}^{T}={\left|
{\left(Q_{1}Q_{2}\right)}_{1}^{\bar{3}}
{\left(\bar{q}_{3}\bar{q}_{4}\right)}_{1}^{3} \right\rangle}_{0}
\delta_{12}^{S}\delta_{34}^{S}, \\ \phi_{1}^{T}\chi_{4}^{T}={\left|
{\left(Q_{1}Q_{2}\right)}_{1}^{6}
{\left(\bar{q}_{3}\bar{q}_{4}\right)}_{0}^{\bar{6}} \right\rangle}_{1}
\delta_{12}^{A}\delta_{34}^{S}, \\ \phi_{2}^{T}\chi_{4}^{T}={\left|
{\left(Q_{1}Q_{2}\right)}_{1}^{\bar{3}}
{\left(\bar{q}_{3}\bar{q}_{4}\right)}_{0}^{3} \right\rangle}_{1}
\delta_{12}^{S}\delta_{34}^{A}, \\ \phi_{1}^{T}\chi_{5}^{T}={\left|
{\left(Q_{1}Q_{2}\right)}_{0}^{6}
{\left(\bar{q}_{3}\bar{q}_{4}\right)}_{1}^{\bar{6}} \right\rangle}_{1}
\delta_{12}^{S}\delta_{34}^{A}, \\ \phi_{2}^{T}\chi_{5}^{T}={\left|
{\left(Q_{1}Q_{2}\right)}_{0}^{\bar{3}}
{\left(\bar{q}_{3}\bar{q}_{4}\right)}_{1}^{3} \right\rangle}_{1}
\delta_{12}^{A}\delta_{34}^{S}, \\ \phi_{1}^{T}\chi_{6}^{T}={\left|
{\left(Q_{1}Q_{2}\right)}_{0}^{6}
{\left(\bar{q}_{3}\bar{q}_{4}\right)}_{0}^{\bar{6}} \right\rangle}_{0}
\delta_{12}^{S}\delta_{34}^{S}, \\ \phi_{2}^{T}\chi_{6}^{T}={\left|
{\left(Q_{1}Q_{2}\right)}_{0}^{\bar{3}}
{\left(\bar{q}_{3}\bar{q}_{4}\right)}_{0}^{3} \right\rangle}_{0}
\delta_{12}^{A}\delta_{34}^{A}, \end{aligned}  \label{colorspinT}
\end{equation}%
where we employ a set of delta notations to reflect the requirement of the
flavor symmetry. The notation $\delta _{34}^{A}=0(\delta _{34}^{S}=0)$ if
two light quarks $q_{3}$ and $q_{4}$ are symmetric (antisymmetric) in flavor
space, and $\delta _{12}^{A}=0$ if two heavy quarks $Q_{1}$ = $Q_{2}$ are
symmetric in flavor space. In all other cases, all these delta notations is
one: $\delta _{34}^{S}=\delta _{34}^{A}=1$, $\delta _{12}^{S}=\delta
_{12}^{A}=1$.

In order for the tetraquarks $T_{QQ}$ in the ground states, we consider two
classes: (i) The isospin $I=0$. In flavor space, light antiquark pair $\bar{n%
}\bar{n}$ is antisymmetric (isosinglet), and two heavy quarks are symmetric
if $QQ=cc$ or $bb$, while they have no certain symmetry if $QQ=bc$; (ii) The
isospin $I=1$. In flavor space, light antiquark pair $\bar{n}\bar{n}$ is
symmetric (isovector), and two heavy quark are symmetric if $QQ=cc$ or $bb$,
and they have no certain symmetry if $QQ=bc$.

With the corresponding wavefunctions, we then use Eq.~(\ref{HCMI}) to
evaluate the CMI matrices for all ground-state configurations. Using
notation $V_{ij}\equiv A_{ij}/\left( m_{i}m_{j}\right) $, the calculated
results of the CMI matrices for the involved color-spin configurations of
the DH tetraquark $QQ\bar{n}\bar{n}$ are shown in Table \ref{tab:CMI} in
details.

\renewcommand{\tabcolsep}{0.30cm} \renewcommand{\arraystretch}{1.1}
\begin{table}[tbh]
\caption{The CMI matrices for the tetraquarks $cc\bar{n}\bar{n}$, $bb\bar{n}%
\bar{n}$ and $bc\bar{n}\bar{n}$ with respective quantum numbers, color-spin
wavefunctions. The parameters $V_{ij}\equiv A_{ij}/\left( m_{i}m_{j}\right) $%
. }\setlength{\abovecaptionskip}{0.5cm}
\begin{tabular}{cccc}
\hline\hline
\textrm{\ System } & $IJ^{P}$ & \textrm{Wave function} & $H_{CMI}$ \\
$cc\bar{n}\bar{n}$ & $01^{+}$ & ($\phi _{2}^{T}\chi _{4}^{T},\phi
_{1}^{T}\chi _{5}^{T}$) & $%
\begin{pmatrix}
\frac{8}{3}V_{cc}-8V_{nn} & -8\sqrt{2}V_{cn} \\
-8\sqrt{2}V_{cn} & 4V_{cc}-\frac{4}{3}V_{nn}%
\end{pmatrix}%
$ \\
& $10^{+}$ & ($\phi _{2}^{T}\chi _{3}^{T},\phi _{1}^{T}\chi _{6}^{T}$) & $%
\begin{pmatrix}
\frac{8}{3}(V_{cc}-4V_{cn}+V_{nn}) & 8\sqrt{6}V_{cn} \\
8\sqrt{6}V_{cn} & 4(V_{cc}+V_{nn})%
\end{pmatrix}%
$ \\
& $11^{+}$ & ($\phi _{2}^{T}\chi _{2}^{T}$) & $\frac{8}{3}%
(V_{cc}-2V_{cn}+V_{nn})$ \\
& $12^{+}$ & ($\phi _{2}^{T}\chi _{1}^{T}$) & $\frac{8}{3}%
(V_{cc}+2V_{cn}+V_{nn})$ \\
$bb\bar{n}\bar{n}$ & $01^{+}$ & ($\phi _{2}^{T}\chi _{4}^{T},\phi
_{1}^{T}\chi _{5}^{T}$) & $%
\begin{pmatrix}
\frac{8}{3}V_{bb}-8V_{nn} & -8\sqrt{2}V_{bn} \\
-8\sqrt{2}A_{bn} & 4V_{bb}-\frac{4}{3}V_{nn}%
\end{pmatrix}%
$ \\
& $10^{+}$ & ($\phi _{2}^{T}\chi _{3}^{T},\phi _{1}^{T}\chi _{6}^{T}$) & $%
\begin{pmatrix}
\frac{8}{3}(V_{bb}-4V_{bn}+V_{nn}) & 8\sqrt{6}V_{bn} \\
8\sqrt{6}V_{bn} & 4(V_{bb}+V_{nn})%
\end{pmatrix}%
$ \\
& $11^{+}$ & ($\phi _{2}^{T}\chi _{2}^{T}$) & $\frac{8}{3}%
(V_{bb}-2V_{bn}+V_{nn})$ \\
& $12^{+}$ & ($\phi _{2}^{T}\chi _{1}^{T}$) & $\frac{8}{3}%
(V_{bb}+2V_{bn}+V_{nn})$ \\
$bc\bar{n}\bar{n}$ & $00^{+}$ & ($\phi _{2}^{T}\chi _{6}^{T},\phi
_{1}^{T}\chi _{3}^{T}$) & $%
\begin{pmatrix}
-8(V_{bc}+V_{nn}) & 4\sqrt{6}(V_{bn}+V_{cn}) \\
4\sqrt{6}(V_{bn}+V_{cn}) & -\frac{4}{3}(V_{bc}+10V_{bn}+10V_{cn}+V_{nn})%
\end{pmatrix}%
$ \\
& $01^{+}$ & ($\phi _{2}^{T}\chi _{4}^{T},\phi _{1}^{T}\chi _{2}^{T},\phi
_{1}^{T}\chi _{5}^{T}$) & $%
\begin{pmatrix}
\frac{8}{3}V_{bc}-8V_{nn} & -8(V_{bn}-V_{cn}) & -4\sqrt{2}(V_{bn}+V_{cn}) \\
-8(V_{bn}-V_{cn}) & -\frac{4}{3}(V_{bc}+5V_{bn}+5V_{cn}+V_{nn}) & -\frac{20%
\sqrt{2}}{3}(V_{bn}-V_{cn}) \\
-4\sqrt{2}(V_{bn}+V_{cn}) & -\frac{20\sqrt{2}}{3}(V_{bn}-V_{cn}) & 4V_{bc}-%
\frac{4}{3}V_{nn}%
\end{pmatrix}%
$ \\
& $02^{+}$ & ($\phi _{1}^{T}\chi _{1}^{T}$) & $-\frac{4}{3}%
(V_{bc}-5V_{bn}-5V_{cn}+V_{nn})$ \\
& $10^{+}$ & ($\phi _{2}^{T}\chi _{3}^{T},\phi _{1}^{T}\chi _{6}^{T}$) & $%
\begin{pmatrix}
\frac{8}{3}(V_{bc}-2V_{bn}-2V_{cn}+V_{nn}) & 4\sqrt{6}(V_{bn}+V_{cn}) \\
4\sqrt{6}(V_{bn}+V_{cn}) & 4(V_{bc}+V_{nn})%
\end{pmatrix}%
$ \\
& $11^{+}$ & ($\phi _{2}^{T}\chi _{2}^{T},\phi _{2}^{T}\chi _{5}^{T},\phi
_{1}^{T}\chi _{4}^{T})$ & $%
\begin{pmatrix}
\frac{8}{3}(V_{bc}-V_{bn}-V_{cn}+V_{nn}) & -\frac{8}{3}\sqrt{2}%
(V_{bn}-V_{cn}) & -8(V_{bn}-V_{cn}) \\
-\frac{8}{3}\sqrt{2}(V_{bn}-V_{cn}) & -8V_{bc}+\frac{8}{3}V_{nn} & -4\sqrt{2}%
(V_{bn}+V_{cn}) \\
-8(V_{bn}-V_{cn}) & -4\sqrt{2}(V_{bn}+V_{cn}) & -\frac{4}{3}(V_{bc}-3V_{nn})%
\end{pmatrix}%
$ \\
& $12^{+}$ & ($\phi _{2}^{T}\chi _{1}^{T}$) & $\frac{8}{3}%
(V_{bc}+V_{bn}+V_{cn}+V_{nn})$ \\ \hline\hline
\end{tabular}%
\label{tab:CMI}
\end{table}

\section{Determination of Parameters}

\label{Parameters}

As a mass formula for hadrons in ground state, Eq. (\ref{Mass}) should be
applicable to the conventional hadrons for which most measured or empirical
data are available and the numbers of quarks in hadrons become less and some
of them may be antiquarks. Further, the binding energy term will be absent
when the hadron considered contains only one heavy quark. In this section,
we demonstrate that the measured mass data from the conventional hadrons are
sufficient to determine three parameters in the formula (\ref{Mass}) and
Eq.~(\ref{HCMI}): $E_{i}$, $\,$the flux-tube energy, $B_{QQ}$, the enhanced
binding energy for the heavy pair $QQ$ in Eq.~(\ref{Mass}), $A_{ij}$, the
CMI coupling between the $i$-the quark and the $j$-th quark in Eq.~(\ref%
{HCMI}).

We note that the binding energy term $B_{QQ}$ is absent when Eq.~(\ref{Mass}%
) applies to mesons and baryons with only one heavy quark. With respect to
the measured masses of the ground-state hadrons with one or two heavy
quarks, we summarize their mass expressions, given by Eq. (\ref{Mass}), in
Table \ref{tab:charmed baryon mass} for the charmed sector and in Table \ref%
{tab:bottom baryon mass} for the bottom sector. There, the masses of the $%
\Xi _{cc}^{\ast }$ are taken from the estimate $M(\Xi _{cc}^{\ast })=M(\Xi
_{cc})+\Delta M(\Xi _{cc})=3728.1$\ MeV, where $\Delta M(\Xi
_{cc})=3/4[M(D^{\ast 0},1^{-})-M{(D^{0},0^{-})}]=106.5$ MeV are obtained by
the heavy quark-diquark symmetry \cite{Mehen:D06,SavageW:D90}. In Table \ref%
{tab:bottom baryon mass}, the masses of the doubly bottom baryons, $M(\Xi
_{bb})=10091.0$\ MeV and\ $M(\Xi _{bb}^{\ast })=10103.0\ $MeV are from the
lattice data \cite{Mohanta:2019mxo}.

\renewcommand{\tabcolsep}{0.46cm} \renewcommand{\arraystretch}{1.1}
\begin{table}[tbh]
\caption{Quark model description of the charmed hadrons, in which $%
m_{u}=m_{d}=m_{n}=230\ \mathrm{MeV}$, $m_{c}=1440\ \mathrm{MeV}$
\protect\cite{Jia:2019bkr}. The masses are in MeV. }
\label{tab:charmed baryon mass}\setlength{\abovecaptionskip}{0.5cm}
\begin{tabular}{cccc}
\hline\hline
Hadron & $J^{P}$ & Mass expression of hadrons & Exp. \cite{Workman:2022ynf}
\\ \hline
$\eta _{c}$ & $0^{-}$ & $2\left( m_{c}+E_{c}\right) +B_{c\bar{c}}-16\frac{%
A_{cc}}{m_{c}m_{c}}$ & $2983.9$ \\
$J/\Psi $ & $1^{-}$ & $2\left( m_{c}+E_{c}\right) +B_{c\bar{c}}+\frac{16}{3}%
\frac{A_{cc}}{m_{c}m_{c}}$ & $3096.9$ \\
$\Lambda _{c}^{+}$ & $\frac{1}{2}^{+}$ & $\left( m_{c}+E_{c}\right) +2\left(
m_{n}+E_{n}\right) -8\frac{A_{nn}}{m_{n}m_{n}}$ & $2286.5$ \\
$\Sigma _{c}$ & $\frac{1}{2}^{+}$ & $\left( m_{c}+E_{c}\right) +2\left(
m_{n}+E_{n}\right) -\frac{32}{3}\frac{A_{cn}}{m_{c}m_{n}}+\frac{8}{3}\frac{%
A_{nn}}{m_{n}m_{n}}$ & $2454.0$ \\
$\Sigma _{c}^{\ast }$ & $\frac{3}{2}^{+}$ & $\left( m_{c}+E_{c}\right)
+2\left( m_{n}+E_{n}\right) +\frac{16}{3}\frac{A_{cn}}{m_{c}m_{n}}+\frac{8}{3%
}\frac{A_{nn}}{m_{n}m_{n}}$ & $2518.4$ \\
$\Xi _{cc}$ & $\frac{1}{2}^{+}$ & $2\left( m_{c}+E_{c}\right) +\left(
m_{n}+E_{n}\right) +B_{cc}-\frac{32}{3}\frac{A_{cn}}{m_{c}m_{n}}+\frac{8}{3}%
\frac{A_{cc}}{m_{c}m_{c}}$ & $3621.6$ \\
$\Xi _{cc}^{\ast }$ & $\frac{3}{2}^{+}$ & $2\left( m_{c}+E_{c}\right)
+\left( m_{n}+E_{n}\right) +B_{cc}+\frac{16}{3}\frac{A_{cn}}{m_{c}m_{n}}+%
\frac{8}{3}\frac{A_{cc}}{m_{c}m_{c}}$ & $3728.1$ \\ \hline\hline
\end{tabular}%
\end{table}

\renewcommand{\tabcolsep}{0.46cm} \renewcommand{\arraystretch}{1.1}
\begin{table}[tbh]
\caption{Quark model description of the bottomed hadrons in ground-state, in
which $m_{u}=m_{d}=m_{n}=230\ \mathrm{MeV}$, $m_{b}=4480\ \mathrm{MeV}$
\protect\cite{Jia:2019bkr}. The masses are in MeV. The data marked with $a$
are from the lattice data \protect\cite{Mohanta:2019mxo}. }
\label{tab:bottom baryon mass}\setlength{\abovecaptionskip}{0.5cm}
\begin{tabular}{cccc}
\hline\hline
Hadron & $J^{P}$ & Mass expression of hadrons & Exp. \cite{Workman:2022ynf}
\\ \hline
$\eta _{b}$ & $0^{-}$ & $2\left( m_{b}+E_{b}\right) +B_{b\bar{b}}-16\frac{%
A_{bb}}{m_{b}m_{b}}$ & $9398.7$ \\
$\Upsilon $ & $1^{-}$ & $2\left( m_{b}+E_{b}\right) +B_{b\bar{b}}+\frac{16}{3}%
\frac{A_{bb}}{m_{b}m_{b}}$ & $9460.3$ \\
$\Lambda _{b}^{0}$ & $\frac{1}{2}^{+}$ & $\left( m_{b}+E_{b}\right) +2\left(
m_{n}+E_{n}\right) -8\frac{A_{nn}}{m_{n}m_{n}}$ & $5169.6$ \\
$\Sigma _{b}$ & $\frac{1}{2}^{+}$ & $\left( m_{b}+E_{b}\right) +2\left(
m_{n}+E_{n}\right) -\frac{32}{3}\frac{A_{bn}}{m_{b}m_{n}}+\frac{8}{3}\frac{%
A_{nn}}{m_{n}m_{n}}$ & $5813.1$ \\
$\Sigma _{b}^{\ast }$ & $\frac{3}{2}^{+}$ & $\left( m_{b}+E_{b}\right)
+2\left( m_{n}+E_{n}\right) +\frac{16}{3}\frac{A_{bn}}{m_{b}m_{n}}+\frac{8}{3%
}\frac{A_{nn}}{m_{n}m_{n}}$ & $5832.5$ \\
$\Xi _{bb}$ & $\frac{1}{2}^{+}$ & $2\left( m_{b}+E_{b}\right) +\left(
m_{n}+E_{n}\right) +B_{bb}-\frac{32}{3}\frac{A_{bn}}{m_{b}m_{n}}+\frac{8}{3}%
\frac{A_{bb}}{m_{b}m_{b}}$ & $10091.0^{a}$ \\
$\Xi _{bb}^{\ast }$ & $\frac{3}{2}^{+}$ & $2\left( m_{b}+E_{b}\right)
+\left( m_{n}+E_{n}\right) +B_{bb}+\frac{16}{3}\frac{A_{bn}}{m_{b}m_{n}}+%
\frac{8}{3}\frac{A_{bb}}{m_{b}m_{b}}$ & $10103.0^{a}$ \\ \hline\hline
\end{tabular}%
\end{table}

\subsection{The coupling paramters $A_{ij}$}

\label{Aij}

We consider first the CMI coupling $A_{ij}$ in Eq.~(\ref{HCMI}) which is
related to the mass splitting between hadrons with same flavor content but
different spin configurations. We evaluate all couplings $A_{ij}$ via the
hyperfine spin splittings of the conventional mesons and baryons provided
that these couplings are approximately same to the CMI couplings for the DH
tetraquarks with color configuration $\bar{3}_{c}\otimes 3_{c}$ in Eq. (\ref%
{ph2}). For the tetraquark with $6_{c}\otimes \bar{6}_{c}$ configuration,
for which chromodynamic is not directly available in the conventional mesons
or baryons associated with the configuration $3_{c}$(or $\bar{3}_{c}$), we
employ the ratios of the color factors of the two configurations to scale
the CMI coupling $A_{ij}$ and $B_{QQ}$ for the configuration $\bar{3}%
_{c}\otimes 3_{c}$ of the tetraquarks to that for configuration $%
6_{c}\otimes \bar{6}_{c}$.

Application of Eq. (\ref{Mass}) to the heavy hadrons of the system $c\bar{c}$%
, $b\bar{b}$, $b\bar{c}$, $ccn$ and $bbn$, as well as the light nulceon
systems of the $N$ and $\Delta $ lead to the mass splitings Eq.~(\ref{Acc})
through Eq.~(\ref{Ann}),

\begin{equation}
M(J/\Psi )-M(\eta _{c})=\frac{64}{3}\frac{A_{cc}}{m_{c}m_{c}},  \label{Acc}
\end{equation}

\begin{equation}
M(\Upsilon )-M(\eta _{b})=\frac{64}{3}\frac{A_{bb}}{m_{b}m_{b}},
\end{equation}

\begin{equation}
M(B_{c}^{+})-M(B_{c}^{\ast +})=\frac{64}{3}\frac{A_{bc}}{m_{b}m_{c}},
\end{equation}

\begin{equation}
M(\Xi _{cc}^{\ast })-M(\Xi _{cc})=16\frac{A_{cn}}{m_{c}m_{n}},
\end{equation}

\begin{equation}
M(\Xi _{bb}^{\ast })-M(\Xi _{bb})=16\frac{A_{bn}}{m_{b}m_{n}},
\end{equation}

\begin{equation}
M(\Delta )-M(N)=16\frac{A_{nn}}{m_{n}m_{n}},  \label{Ann}
\end{equation}%
where masses of the nucleons have the form, in terms of Eq. (\ref{Mass}),
\begin{eqnarray}
M(N) &=&3\left( m_{n}+E_{n}\right) -\frac{8A_{nn}}{m_{n}m_{n}}=939.6\ \text{%
MeV,}  \label{MN} \\
M(\Delta ) &=&3\left( m_{n}+E_{n}\right) +\frac{8A_{nn}}{m_{n}m_{n}}=1232.0\
\text{MeV.}  \label{MD}
\end{eqnarray}%
Given the measured masses \cite{Workman:2022ynf} of the hadrons $J/\Psi $,
the $\eta _{c}$, the $\Upsilon $, $\eta _{b}$ and other heavy baryons in Eq.
(\ref{Acc}) through Eq. (\ref{Ann}), as well as the nucleaon masses given in
Eq. (\ref{MN}) and Eq. (\ref{MD}), the $M(B_{c}^{\ast +})=6332.0\ $MeV in
Ref \cite{Ebert:2002pp}, one can use Eq.~(\ref{Acc}) through Eq.~(\ref{Ann})
to determine the CMI coupling parameters $A_{cc}$, $A_{bb}$, $A_{bc}$, $%
A_{cn}$, $A_{bn}$, $A_{nn}$. The obtained results for these couplings are
collected in Table \ref{tab:coefficients CMI}.

\renewcommand{\tabcolsep}{0.46cm} \renewcommand{\arraystretch}{1.1}
\begin{table}[tbh]
\caption{The extracted coupling parameters for the CMI}\setlength{%
\abovecaptionskip}{0.5cm}
\begin{tabular}{ccccccc}
\hline\hline
Coupling & $ij=cc$ & $ij=cn$ & $ij=bb$ & $ij=bn$ & $ij=bc$ & $ij=nn$ \\
\hline
$A_{ij}$ (GeV$^{3}$) & 0.01099 & 0.00221 & 0.05800 & 0.007723 & 0.01742 &
0.00970 \\ \hline
$\frac{A_{ij}}{m_{i}m_{j}}$ (MeV) & 5.30 & 6.66 & 2.89 & 0.75 & 2.70 & 18.34
\\ \hline\hline
\end{tabular}%
\label{tab:coefficients CMI}
\end{table}

\subsection{The binding energy $B_{QQ}$($Q=c$, $b$)}

\label{BQQ}

\begin{figure}[th]
\begin{center}
\includegraphics[width=0.77\textwidth]
{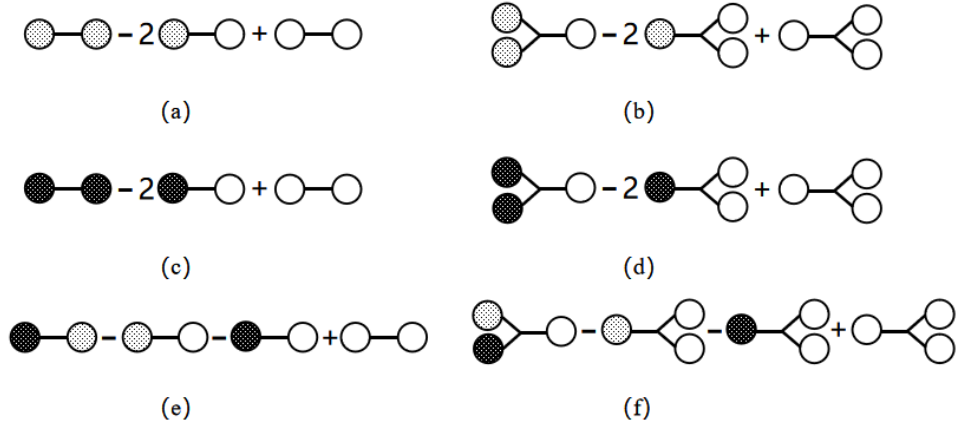}
\end{center}
\caption{The binding energies $B_{Q\bar{Q}}$ and $B_{QQ}$ between one heavy
and one antiheavy quarks and between two heavy quarks are evaluted by the
sum of the mean masses of the first and the last hadrons minus that of
others. The black (grey) solid circle stands for the bottom (charm) quark $b$(%
$c$). The hollow circle stands for the nonstrange ($u,d$) quarks.}
\label{Fig1}
\end{figure}

To extract the binding energy $B_{QQ}$($Q=c$, $b$) for two heavy quarks $QQ$
and $B_{Q\bar{Q}}$ for a heavy-antiheavy pair $Q\bar{Q}$ from heavy
conventional hadrons (mesons, baryons), we utilize a set of simple relations
relating the measured masses of the established hadrons. The main idea of
the relations for the $B_{Q\bar{Q}}$ is illustrated in FIG.~\ref{Fig1}. Take
FIG.~\ref{Fig1}(a) for instance, it shows that the binding energy ($B_{c\bar{%
c}}$) between the charm quark $c$ and the anticharm quark $\bar{c}$ is
measured to be the sum of mean masses of the $c\bar{c}$ system ($M_{c\bar{c}%
} $) and the light (nonstrange) meson $n\bar{n}$ systems ($M_{n\bar{n}}$),
subtracted the mean mass of the singly charmed mesons ($M_{c\bar{n}}$)
twice, which conceals all charm(anticharm) quark masses and that of the
light quarks, left only with binding energy. The procedures for $B_{b\bar{b}%
} $ in FIG.~\ref{Fig1}(c) and for $B_{b\bar{c}}$ in FIG.~\ref{Fig1}(e) are
similar: the pair $c\bar{c}$ and the $c\bar{n}$ replaced by the $b\bar{b}$
and the $b\bar{n}$, or by the pair $b\bar{c}$ and both of the $c\bar{n}$ and
the $b\bar{n}$, respectively. In the cases of the binding $B_{QQ}$ between
pair $QQ$, it is measured to be the sum of mean masses of the DH baryons ($%
M_{QQn}$) and the light nucleons ($M_{nnn}$), subtracted the mean mass of
singly heavy baryons ($M_{Qnn}$) twice, which conceals all heavy quark
masses and that of the light quarks. This relation is shown in FIG.~\ref%
{Fig1}(b) for the system $ccn$, FIG.~\ref{Fig1}(d) for the $bbn$ and FIG.~%
\ref{Fig1}(f) for the $bcn$, where the black(grey) solid cirlce stands for
the bottom(charm) quark while the hollow circle stands for the light quarks.
Corresponding to these subfigures, the relations are listed in Table \ref%
{tab:binding energy} collectively.

\renewcommand{\tabcolsep}{0.46cm} \renewcommand{\arraystretch}{1.1}
\begin{table}[tbh]
\caption{The binding energies between one heavy quark and another heavy
quark or antiquark and their values computed via the relations in the first
column.}
\label{tab:binding energy}\setlength{\abovecaptionskip}{0.5cm}
\begin{tabular}{cc}
\hline\hline
The binding energy & Value (MeV) \\ \hline
$B_{c\bar{c}}=M_{c\bar{c}}-2M_{c\bar{n}}+M_{n\bar{n}}$ & $-257.9$ \\
$B_{b\bar{b}}=M_{b\bar{b}}-2M_{b\bar{n}}+M_{n\bar{n}}$ & $-562.0$ \\
$B_{b\bar{c}}=M_{b\bar{c}}-M_{b\bar{n}}-M_{c\bar{n}}+M_{n\bar{n}}$ & $-349.1$
\\
$B_{cc}=M_{ccn}-2M_{cnn}+M_{nnn}$ & $-166.8$ \\
$B_{bb}=M_{bbn}-2M_{bnn}+M_{nnn}$ & $-418.6$ \\
$B_{bc}=M_{bcn}-M_{bnn}-M_{cnn}+M_{nnn}$ & $-217.5$ \\ \hline\hline
\end{tabular}%
\end{table}

Given the relations in Table \ref{tab:binding energy}, one can estimate $%
B_{QQ}$ (and $B_{Q\bar{Q}}$) using the measured (mean) masses \cite%
{Workman:2022ynf} of the observed mesons and baryons involved, with the
results listed in the second column of Table \ref{tab:binding energy}. One
exception is $B_{bc}$ for which the masses of the baryons $\Xi _{bc}$ and $%
\Xi _{bc}^{\ast }$ are from the lattice compution, $M(\Xi _{bc})=6943.0$
MeV, $M(\Xi _{bc}^{\ast })=6985.0\ $MeV \cite{Brown:2014ena}. Notice that in
Table \ref{tab:binding energy} the notation $M$ with quark subscript
respresents the mean masses of the hadrons in the sense that the CMI are
minimized. All mean masses for the hadrons involved are evaluated from the
measured data of the hadrons and shown in Table \ref{tab:spin-weighted mass}
explicitly, with exception for the pion mesons with $M(\pi )=154.0\ $MeV
predicted by relativistic quark model \cite{Ebert:2017els}. We choose this
value which is slightly larger than measured data ($140$ MeV) for the pion
since the pion is believed to be suppressed in mass normally by chiral
symmetry(the Nambu-Goldstone mechanisim) \cite{ChengTP:1985}. For $B_{Q\bar{Q}}$, our values
are in consistent with that predicted by Ref. \cite{Karliner:2014gca}, for
instance, our value $B_{c\bar{c}}=-257.9\ $MeV is close to the prediction $%
B_{c\bar{c}}=-258.0\ $MeV in Ref. \cite{Karliner:2014gca}.

\renewcommand{\tabcolsep}{0.25cm} \renewcommand{\arraystretch}{1.1}
\begin{table}[tbh]
\caption{The expressions and their values of the mean masses of hadrons in
their ground states.}\setlength{\abovecaptionskip}{0.5cm}
\begin{tabular}{cccc}
\hline\hline
The spin-averaged mass & Value (MeV) & The spin-averaged mass & Value (MeV)
\\ \hline
$M_{c\bar{c}}=[M(\eta _{c})+3M(J/\Psi )]/4$ & 3068.7 & $M_{ccn}=[2M(\Xi
_{cc})+4M(\Xi _{cc}^{\ast })]/6$ & 3692.6 \\
$M_{c\bar{n}}=[M(D^{\pm })+3M(D^{\ast \pm })/4$ & 1973.2 & $%
M_{cnn}=[2M(\Sigma _{c})+4M(\Sigma _{c}^{\ast })]/6$ & 2496.9 \\
$M_{n\bar{n}}=[M(\pi )+3M(\rho )]/4$ & 620.0 & $M_{nnn}=[2M(n)+4M(\Delta )]/6
$ & 1134.5 \\
$M_{b\bar{b}}=[M(\eta _{b})+3M(\Upsilon )]/4$ & 9444.9 & $M_{bbn}=[2M(\Xi
_{bb})+4M(\Xi _{bb}^{\ast })]/6$ & 10099.0 \\
$M_{b\bar{n}}=[M(B)+3M(B^{\ast })]/4$ & 5313.4 & $M_{bnn}=[2M(\Sigma
_{b})+4M(\Sigma _{b}^{\ast })]/6$ & 5826.1 \\
$M_{c\bar{b}}=[M(B_{c}^{+})+3M(B_{c}^{\ast +})]/4$ & 6317.6 & $%
M_{bcn}=[2M(\Xi _{bc})+4m(\Xi _{bc}^{\ast })]/6$ & 6971.0 \\ \hline\hline
\end{tabular}%
\label{tab:spin-weighted mass}
\end{table}

\subsection{The flux energy $E_{i}$}

We assume in this work that the DH baryons and the singly heavy baryons are
similar in static dynamics to the DH tetraquarks with the same heavy flavor.
In this approximation, one can estimate the flux-tube energy $E_{i}$ tied to the
i-th quark in Eq. (\ref{Mass}). As the values of $E_{i}$ may depend on the
flavor content of the DH baryons $Q_{1}Q_{2}n$ ($Q_{1},Q_{2}=c,b$), we
consider the three cases:

(i) The doubly charmed baryons $\Xi _{cc}=ccn$: In this situation, one can
apply Eq. (\ref{Mass}) to the $\Xi _{cc}$ and $\Xi _{cc}^{\ast }$ and find
the mean mass $M_{ccn}$ for them to be,
\begin{equation}
2\left( m_{c}+E_{c}\right) +\left( m_{n}+E_{n}\right) +B_{cc}+\frac{8}{3}%
\frac{A_{cc}}{m_{c}m_{c}}=M_{ccn}\mathrm{.}  \label{Xcc}
\end{equation}%
Further, the same method applied to the charmed baryons $\Sigma _{c}$, $%
\Sigma _{c}^{\ast }$ and $\Lambda _{c}^{+}$ leads to the relation for $M_{cnn}$,
\begin{equation}
\left( m_{c}+E_{c}\right) +2\left( m_{n}+E_{n}\right) =M_{cnn}\mathrm{.}
\label{SL}
\end{equation}%
With the values $M_{ccn}=3692.6\ $MeV and $M_{cnn}=2444.3\ $MeV in Table VI$%
\mathrm{,}$ Eqs. (\ref{Xcc}) and (\ref{SL}) allow one to solve for two
parameters $E_{c}$ and $E_{n}$. The results are $E_{c}=308.7$ MeV, $%
E_{n}=117.8$ MeV.

(ii) The doubly bottom baryons $\Xi _{bb}=bbn$: For this case, one can apply
Eq. (\ref{Mass}) to the $\Xi _{bb}$ and $\Xi _{bb}^{\ast }$ and thereby
find,
\begin{equation}
\ 2\left( m_{b}+E_{b}\right) +\left( m_{n}+E_{n}\right) +B_{bb}+\frac{8}{3}%
\frac{A_{bb}}{m_{b}m_{b}}=M_{bbn}\text{.}  \label{Xbb}
\end{equation}%
Similar method applied to the baryons $\Sigma _{b}$, $\Sigma _{b}^{\ast }$ and $\Lambda _{b}^{0}$ give

\begin{equation}
\left( m_{b}+E_{b}\right) +2\left( m_{n}+E_{n}\right) =M_{bnn}\text{. }
\label{E2}
\end{equation}%
Combining $M_{bbn}=10099.0\ $MeV and $M_{bnn}=5826.1\ $MeV given in Table
VI, Eqs. (\ref{Xbb}) and (\ref{E2}) lead to the solutions: $E_{b}=601.7$
MeV, $E_{n}=E_{n}^{\ast }\equiv 116.3$\ MeV. Here, the superscript star is
used to denote the flux-tube energy $E_{n}$ in the singly bottom baryons.

(iii) The bottom-charmed baryons $\Xi _{bc}=bcn$: In this case, application
of Eq. (\ref{Mass}) to the $\Xi _{bc}$ and $\Xi _{bc}^{\ast }$ leads to the
following relation for $M_{bcn}$,
\begin{equation}
\left( m_{b}+E_{b}\right) +\left( m_{c}+E_{c}\right) +\left(
m_{n}+E_{n}\right) +B_{cb}+\frac{8}{3}\frac{A_{bc}}{m_{c}m_{b}}=M_{bcn}\text{%
,}  \label{Xbc}
\end{equation}%
from which using $M_{bcn}=6971.0$\ MeV in Table VI one can solve $%
E_{c}^{\ast }$ $=312.7$ MeV, $E_{b}^{\ast }$ $=602.8$\ MeV and $%
E_{n}=E_{n}^{\ast \ast }=115.8$\ MeV. Here, the superscript star in $%
E_{c,b}^{\ast }$ are used to denote the values of $E_{c,b}$ in the
bottom-charmed baryons, and the double star stands for the values of $E_{n}$
in the bottom-charmed baryons.

One sees that the values of the $E_{b}$ in the bottom baryons ($601.7$ MeV)
and the bottom-charmed baryons ($602.8$\ MeV) are almost same while the
values of the $E_{c}$ in the charmed baryons ($308.7$ MeV) and the
bottom-charmed baryons ($312.7$ MeV) differ slightly (with uncertainty $4$\
MeV). The values of the flux-tube energies $E_{n}$ of the up and down quarks
are nearly identical ($E_{n}=117.8$ MeV, $E_{n}^{\ast }=116.3$ MeV and $%
E_{n}^{\ast \ast }=115.8$\ MeV). In this work, we choose to use the
respective values of the flux-tube energies $E_{c,b}$ and $E_{n}$ for the
three cases above though they are close in three cases.

\section{Masses of $T_{cc}$ and other DH tetraquarks}

\label{results}

The binding energies $B_{QQ}=B_{QQ}[\bar{3}_{c}\otimes 3_{c}]$ discussed in
subsection~\ref{BQQ} correspond to the color configuration $\bar{3}_{c}\otimes
3_{c}$ of the baryons we consider. According to the ratio $-1:2$ of the
color factors for the configurations $6_{c}\otimes \bar{6}_{c}$ and $\bar{3}_{c}%
\otimes 3_{c}$, the binding energy for the pair $QQ$ in the tetraquarks in $%
6_{c}\otimes \bar{6}_{c}$ should be given by $B_{QQ}[6_{c}\otimes \bar{6}%
_{c}]=-B_{QQ}/2$ \cite{LuoCLL:2017}, as we shall apply in this work.

Since $B_{QQ}$ form a diagonal matrix in Eq. (\ref{Mass}), the possible
mixing of the color-spin states stems only from the the CMI hamiltonian $%
H_{CMI}$ in Table \ref{tab:CMI}. Given the values of three parameters, $%
A_{ij}$, $B_{QQ}$ and $E_{c,b,n}$, determined in Section~\ref{Parameters},
one can compute the ground-state masses of the doubly heavy tetraquarks $%
T_{cc}$, $T_{bb}$ and $T_{bc}$ via diagonalizing the Sakharov--Zeldovich
Hamiltonian (\ref{Mass}) with the CMI (\ref{HCMI}). The numerical
diagonalization of the matrix $B_{QQ}+H_{CMI}$ are carried out and
illustrated explicitly in the Table \ref{tab:NumR} for the tetraquarks $%
T_{cc}$, $T_{bb}$ and $T_{bc}$ with various quantum numbers($I,J^{P}$), with
the obtained eigenvalues, eigenvectors and the corresponding masses given by
Eq. (\ref{Mass}) shown.\renewcommand{\tabcolsep}{0.22cm} \renewcommand{%
\arraystretch}{1.1}
\begin{table}[tbh]
\caption{Computed $H$ matrices of the nonstrange DH tetraquarks $T_{QQ}=QQ%
\bar{n}\bar{n}$($QQ=cc,bb,bc$) and ensuing diagonalization with the mixed
weights and tetraquark masses (MeV) obtained. Three sets of $E_{i}$ slightly
different (see text) are employed: $E_{c}=308.7$ MeV, $E_{n}=117.8$ MeV; $%
E_{b}=601.7$ MeV, $E_{n}^{\ast }=116.3$\ MeV; $E_{c}^{\ast }=312.7$ MeV, $%
E_{b}^{\ast }=602.8$\ MeV and $E_{n}^{\ast \ast }=115.8$\ MeV.}\textrm{%
\setlength{\abovecaptionskip}{0.5cm}
\begin{tabular}{cccccc}
\hline\hline
Systems\textrm{\ } & $IJ^{P}$ & $H$ $($MeV$)$ & \textrm{Eigenvalues (MeV)} &
\textrm{Eigenvectors} & \textrm{M (MeV)} \\ \hline
$cc\bar{n}\bar{n}$ & $01^{+}$ & $%
\begin{pmatrix}
-299.4 & -75.4 \\
-75.4 & 80.1%
\end{pmatrix}%
$ & $%
\begin{bmatrix}
-313.8 \\
94.6%
\end{bmatrix}%
$ & $%
\begin{bmatrix}
(-0.98,-0.19) \\
(0.19,-0.98)%
\end{bmatrix}%
$ & $%
\begin{bmatrix}
3879.2 \\
4287.6%
\end{bmatrix}%
$ \\
$cc\bar{n}\bar{n}$ & $10^{+}$ & $%
\begin{pmatrix}
-174.8 & 130.5 \\
130.5 & 178.0%
\end{pmatrix}%
$ & $%
\begin{bmatrix}
221.0 \\
-217.8%
\end{bmatrix}%
$ & $%
\begin{bmatrix}
(0.31,0.95) \\
(-0.95,0.31)%
\end{bmatrix}%
$ & $%
\begin{bmatrix}
4414.0 \\
3975.2%
\end{bmatrix}%
$ \\
& $11^{+}$ & -139.3 & -139.3 & 1.00 & 4053.7 \\
& $12^{+}$ & -68.3 & -68.3 & 1.00 & 4124.7 \\
$bb\bar{n}\bar{n}$ & $01^{+}$ & $%
\begin{pmatrix}
-557.6 & -8.5 \\
-8.5 & 196.4%
\end{pmatrix}%
$ & $%
\begin{bmatrix}
-557.7 \\
196.5%
\end{bmatrix}%
$ & $%
\begin{bmatrix}
(-1.00,-0.01) \\
(0.01,-1.00)%
\end{bmatrix}%
$ & $%
\begin{bmatrix}
10298.3 \\
11052.5%
\end{bmatrix}%
$ \\
$bb\bar{n}\bar{n}$ & $10^{+}$ & $%
\begin{pmatrix}
-370.0 & 14.7 \\
14.7 & 294.2%
\end{pmatrix}%
$ & $%
\begin{bmatrix}
-370.3 \\
-294.5%
\end{bmatrix}%
$ & $%
\begin{bmatrix}
(-1.00,0.02) \\
(-0.02,-1.00)%
\end{bmatrix}%
$ & $%
\begin{bmatrix}
10485.7 \\
11150.5%
\end{bmatrix}%
$ \\
& $11^{+}$ & -366.0 & -366.0 & 1.00 & 10490.0 \\
& $12^{+}$ & -358.0 & -358.0 & 1.00 & 10498.0 \\
$bc\bar{n}\bar{n}$ & $00^{+}$ & $%
\begin{pmatrix}
-385.8 & 72.6 \\
72.6 & -18.1%
\end{pmatrix}%
$ & $%
\begin{bmatrix}
-399.6 \\
-4.3%
\end{bmatrix}%
$ & $%
\begin{bmatrix}
(-0.98,0.19) \\
(-0.19,-0.98)%
\end{bmatrix}%
$ & $%
\begin{bmatrix}
7127.5 \\
7522.8%
\end{bmatrix}%
$ \\
& $01^{+}$ & $%
\begin{pmatrix}
-357.0 & 47.3 & -42.0 \\
47.3 & 30.3 & 55.7 \\
-41.9 & 55.7 & 95.1%
\end{pmatrix}%
$ & $%
\begin{bmatrix}
-367.8 \\
127.7 \\
9.4%
\end{bmatrix}%
$ & $%
\begin{bmatrix}
(0.99,-0.13,0.11) \\
(-0.03,0.49,0.87) \\
(-0.17,-0.86,0.48)%
\end{bmatrix}%
$ & $%
\begin{bmatrix}
7159.3 \\
7654.8 \\
7536.5%
\end{bmatrix}%
$ \\
& $02^{+}$ & 130.1 & 130.1 & 1.00 & 7657.2 \\
$bc\bar{n}\bar{n}$ & $10^{+}$ & $%
\begin{pmatrix}
-200.9 & 72.6 \\
72.6 & 192.9%
\end{pmatrix}%
$ & $%
\begin{bmatrix}
-213.8 \\
205.9%
\end{bmatrix}%
$ & $%
\begin{bmatrix}
(-0.98,0.18) \\
(-0.18,-0.98)%
\end{bmatrix}%
$ & $%
\begin{bmatrix}
7313.3 \\
7733.0%
\end{bmatrix}%
$ \\
& $11^{+}$ & $%
\begin{pmatrix}
-118.1 & 22.3 & 47.3 \\
22.3 & -190.2 & -42.0 \\
47.3 & -41.9 & 178.5%
\end{pmatrix}%
$ & $%
\begin{bmatrix}
-204.2 \\
189.7 \\
-115.3%
\end{bmatrix}%
$ & $%
\begin{bmatrix}
(0.32,-0.94,-0.14) \\
(-0.14,-0.10,0.98) \\
(0.94,0.34,-0.10)%
\end{bmatrix}%
$ & $%
\begin{bmatrix}
7322.9 \\
7716.8 \\
7411.8%
\end{bmatrix}%
$ \\
& $12^{+}$ & -141.6 & -141.6 & 1.00 & 7385.5 \\ \hline\hline
\end{tabular}%
}
\label{tab:NumR}
\end{table}

\renewcommand{\tabcolsep}{0.20cm} \renewcommand{\arraystretch}{1.1}
\begin{table}[tbh]
\caption{Our results ($M$) and other computations for masses of the DH
tetraquarks $T_{QQ}$. The lowest threshold ($T$) of two heavy-mesons($Q\bar{q%
}$) for the tetraquarks considered and $\Lambda =M-T$ are also given. All in
MeV. }
\label{tab:comp1}\textrm{\setlength{\abovecaptionskip}{0.5cm}
\begin{tabular}{ccccccccccc}
\hline\hline
\textrm{\ System } & $IJ^{P}$ & $M$ & $T$ & $\Lambda $ & \cite%
{Eichten:2017ffp} & \cite{Song:2023ufz} & \cite{Weng:2021hje} & \cite%
{Wang:2017dtg} & \cite{Gao:2020ogo} &  \\ \hline
$cc\bar{n}\bar{n}$ & $01^{+}$ & 3879.2 & 3875.9($DD^{\ast }$) & 3.3 & 3978.0
& 3997.0 & 3868.7 &  & 3877.0 &  \\
&  & 4287.6 & 4017.2($D^{\ast }D^{\ast }$) & 270.4 &  &  & 4230.8 &  &  &
\\
& $10^{+}$ & 3975.2 & 3875.9($DD^{\ast }$) & 99.3 & 4146.0 & 4163.0 & 3969.2
& 3870.0 &  &  \\
&  & 4414.0 & 4017.2($D^{\ast }D^{\ast }$) & 396.8 &  &  & 4364.9 &  &  &
\\
& $11^{+}$ & 4053.7 & 4017.2($D^{\ast }D^{\ast }$) & 36.5 & 4167.0 & 4185.0
& 4053.2 & 3900.0 & 4021.0 &  \\
& $12^{+}$ & 4124.7 & 4017.2($D^{\ast }D^{\ast }$) & 107.5 & 4210.0 & 4229.0
& 4123.8 & 3950.0 &  &  \\
$bb\bar{n}\bar{n}$ & $01^{+}$ & 10298.3 & 10604.2($BB^{\ast }$) & -305.9 &
10482.0 & 10530.0 & 10390.9 &  & 10353.0 &  \\
&  & 11052.5 & 10649.4($B^{\ast }B^{\ast }$) & 403.1 &  &  & 10950.3 &  &  &
\\
& $10^{+}$ & 10485.7 & 10559.0($BB$) & -73.3 & 10674.0 & 10726.0 & 10569.3 &
&  &  \\
&  & 11150.5 & 10649.4($B^{\ast }B^{\ast }$) & 501.1 &  &  & 11054.6 &  &  &
\\
& $11^{+}$ & 10490.0 & 10604.2($BB^{\ast }$) & -114.2 & 10681.0 & 10733.0 &
10584.2 &  & 10403.0 &  \\
& $12^{+}$ & 10498.0 & 10649.4($B^{\ast }B^{\ast }$) & -151.4 & 10695.0 &
10747.0 & 10606.8 &  &  &  \\
$bc\bar{n}\bar{n}$ & $00^{+}$ & 7127.5 & 7146.8($BD$) & -19.3 & 7229.0 &
7268.0 & 7124.6 &  &  &  \\
&  & 7522.8 & 7333.3($B^{\ast }D^{\ast }$) & 189.5 &  &  & 7459.0 &  &  &
\\
& $01^{+}$ & 7159.3 & 7192.0($B^{\ast }D$) & -32.7 & 7272.0 & 7269.0 & 7158.0
&  &  &  \\
&  & 7536.5 & 7333.3($B^{\ast }D^{\ast }$) & 203.2 &  &  & 7482.4 &  &  &
\\
&  & 7654.8 & 7333.3($B^{\ast }D^{\ast }$) & 321.5 &  &  & 7584.9 &  &  &
\\
& $02^{+}$ & 7657.2 & 7333.3($B^{\ast }D^{\ast }$) & 323.9 &  &  & 7610.3 &
&  &  \\
& $10^{+}$ & 7313.3 & 7288.1($BD^{\ast }$) & 25.2 & 7461.0 & 7458.0 & 7305.6
&  &  &  \\
&  & 7733.0 & 7333.3($B^{\ast }D^{\ast }$) & 399.7 &  &  & 7684.7 &  &  &
\\
& $11^{+}$ & 7322.9 & 7288.1($BD^{\ast }$) & 34.8 & 7472.0 & 7469.0 & 7322.5
&  &  &  \\
&  & 7411.8 & 7333.3($B^{\ast }D^{\ast }$) & 78.5 &  & 7478.0 & 7367.3 &  &
&  \\
&  & 7716.8 & 7333.3($B^{\ast }D^{\ast }$) & 383.5 &  &  & 7665.1 &  &  &
\\
& $12^{+}$ & 7385.5 & 7333.3($B^{\ast }D^{\ast }$) & 52.2 & 7493.0 & 7490.0
& 7396.0 &  &  &  \\ \hline\hline
\end{tabular}%
}
\end{table}

We take the $cc\bar{n}\bar{n}$ tetraquark with $IJ^{P}=01^{+}$ as an
example. As shown in Table \ref{tab:CMI}, the allowed color-spin basis are $%
(\phi _{2}^{T}\chi _{4}^{T},\phi _{1}^{T}\chi _{5}^{T})$, and the CMI matrix
(in MeV)$\allowbreak $ $\allowbreak $ $\allowbreak $
\begin{equation}
\langle H_{CMI}\rangle =\left[
\begin{array}{cc}
\frac{8}{3}V_{cc}-8V_{nn} & -8\sqrt{2}V_{cn} \\
-8\sqrt{2}V_{cn} & 4V_{cc}-\frac{4}{3}V_{nn}%
\end{array}%
\right] =-\left[
\begin{array}{cc}
132.\,\allowbreak 59 & 75.\,\allowbreak 35 \\
75.\,\allowbreak 35 & 3.25%
\end{array}%
\right] ,  \label{hcmi}
\end{equation}%
with $V_{ij}\equiv A_{ij}/\left( m_{i}m_{j}\right) $. The corresponding
matrix of the binding energy (MeV) is
\begin{equation}
B_{QQ}=%
\begin{pmatrix}
B_{cc} & 0 \\
0 & -B_{cc}/2%
\end{pmatrix}%
=\left[
\begin{array}{cc}
-166.8 & 0 \\
0 & 83.4%
\end{array}%
\right] ,  \label{Bcc}
\end{equation}%
for $QQ=cc$, which leads to a sum of two matrices in Eqs. (\ref{hcmi}) and (%
\ref{Bcc}),
\begin{equation}
H=B_{QQ}+\langle H_{CMI}\rangle =\left[
\begin{array}{cc}
-299.\,\allowbreak 4 & -75.\,\allowbreak 4 \\
-75.\,\allowbreak 4 & 80.1%
\end{array}%
\right] .  \label{H}
\end{equation}

The diagonalization of the matrix $H$ in Eq. (\ref{H}) gives rise to the
eigenvalues and the respective eigenvectors,
\begin{equation}
\text{Eig(}H\text{)=}\left[
\begin{array}{c}
-313.8 \\
94.6%
\end{array}%
\right] \text{, Eigv(}H\text{)}_{v}\text{=}\left[
\begin{array}{c}
(-0.98,-0.19) \\
(0.19,-0.98)%
\end{array}%
\right] ,  \label{Eig}
\end{equation}%
by which one can use Sakharov--Zeldovich formula (\ref{Mass}) where $%
M=2(m_{c}+E_{n})+$Eig($H$) to find the ground-state masses of the tetraquark
$T_{cc}$ with $IJ^{P}=01^{+}$,
\begin{eqnarray}
M(T_{cc},01^{+}) &=&3879.2\text{ MeV, in }\bar{3}\otimes 3_{c}\text{ (}96\%%
\text{),}  \label{MTccA} \\
M(T_{cc},01^{+}) &=&4287.6\text{ MeV, in }6\otimes \bar{6}_{c}\text{ (}96\%%
\text{),}  \label{MTccB}
\end{eqnarray}%
with the chromomagnetic-mixing probability $0.98^{2}=\allowbreak 96\%$.

The first value of the $T_{cc}$ masses here is quite close to the $D^{\ast
+}D^{0}$ threshold $M(D^{\ast +}D^{0})=3875.1\pm 0.1$ MeV \cite%
{Workman:2022ynf}, being in agreement with the observed mass of about $%
3875.7\pm 0.05$ MeV of the $J^{P}=1^{+}$ tetraquark $T_{cc}^{+}$($3875$) \cite%
{LHCb:2021vvq,LHCb:2021auc} discovered LHCb at CERN. Notice further that the
LHCb-reported tetraquark $T_{cc}^{+}(3875)$ is very narrow (width$=410\pm
165 $ keV), one can conclude that the lowest state of the doubly-charmed
tetraquark is the $J^{P}=1^{+}$ tetraquark $T_{cc}^{+}$($3875$) that was
observed by LHCb, which is the resonance state of the tetraquark system $cc%
\bar{u}\bar{d}$ with main color configuration of the $\bar{3}_{c}\otimes 3_{c}$.
One infers, by Eq. (\ref{MTccB}), that there exists a nontrivial color
partner $T_{cc}^{+}$($cc\bar{u}\bar{d}$, $6_{c}\otimes \bar{6}_{c}$) of the
tetraquark $T_{cc}^{+}$($3875$), being a fully new tetraquark $cc\bar{u}\bar{%
d}$ with color configuration $6_{c}\otimes \bar{6}_{c}$, and with the same $%
IJ^{P}=01^{+}$ but the mass about $4288$ MeV. Differing only in color
representation from the $T_{cc}^{+}$($3875$), we hope that a further
experimental search similar to the LHCb experiment \cite{LHCb:2021vvq} in the
some final states (e.g., the $D^{0}D^{0}\pi ^{+}$) with inviariant mass
about $408$ MeV higher than the $T_{cc}^{+}$($3875$) can test our
prediction. The numerical diagonalization of the matrix $H=B_{QQ}+H_{CMI}$
are also performed for the $I=1$ tetraquark $T_{cc}$ with $J^{P}=0^{+},1^{+}$
and $2^{+}$, resulting in small color-spin mixing for $J^{P}=0^{+}$ state,
as illustrated in the Table \ref{tab:NumR}.

Further computations are performed similarly for other DH tetraquarks $T_{bb}
$ and $T_{bc}$ with $J^{P}=0^{+},1^{+},2^{+}$ and $I=0,1$, as shown in Table %
\ref{tab:NumR}, resulting in a similar mass order and configurations, except
that the inverse mass order ($4414.0,3975.2$) MeV for the ($T_{cc}^{+}(\bar{3%
}_{c}\otimes 3_{c}),T_{cc}^{+}$($6_{c}\otimes \bar{6}_{c}$)) are replaced by the
normal order ($10485.7,11150.5$) MeV for the ($T_{bb}^{+}(\bar{3}_{c}\otimes
3_{c}),T_{bb}^{+}$($6_{c}\otimes \bar{6}_{c}$)). The $I=1$ tetraquark $T_{bb}$
with $J^{P}=1^{+}$ and that with $J^{P}=2^{+}$ becomes nearly degenerated ($%
10490.0$ MeV, $10498.0$ MeV), in constrast with the splitted masses ($%
4053.7\ $MeV, $4124.7$ MeV) of the $I=1$ tetraquark $T_{cc}$ with $%
J^{P}=1^{+}$ and $2^{+}$. In the case of the bottom-charmed tetraquarks $%
T_{bc}$, the chromomagnetical mixing occurs between two color reps, for the $%
J^{P}=0^{+}$ and among three color-spin states for the $J^{P}=1^{+}$ case,
deeply for the the $IJ^{P}=01^{+}$. Generally, the chromomagnectic-mixing
enters more or less for the $IJ^{P}=01^{+}$ and $IJ^{P}=1(0^{+},1^{+})$
states of all DH tetraquarks $T_{QQ}$, for which the novel $6_{c}\otimes \bar{6}%
_{c}$ color state of the tetraquarks occupies the higher mass levels.

In Table \ref{tab:comp1}, our results for the DH tetraquark masses ($M$) are
compared to other computations via the quark model \cite%
{Eichten:2017ffp,Song:2023ufz}, the chromomagnetic models \cite{Weng:2021hje}
and the QCD sum rules \cite{Wang:2017dtg,Gao:2020ogo}. In the Table, we also
list the lowest thresholds ($T$) of two heavy mesons ($Q\bar{q}$) as
decaying final states of $T_{QQ}$ and the respective mass difference $%
\Lambda =M-T$.

\section{Conclusions and remarks}

\label{Conclusions}

Inspired by recent experimental progress of the tetraquarks, we
systematically study the nonstrange doubly-heavy tetraquarks $T_{cc}$, $%
T_{bb}$ and $T_{bc}$ in the chromomagnetic interaction model with unified
quark masses and the flux-tube correction incorporated. A binding energy $%
B_{QQ}$ between heavy-heavy quarks is introduced and estimated via a simple
relation in terms of the measured masses of baryons, by which we extract the
flux-tube conrrections $E_{i}$ associated with $i$-th quark in doubly-heavy
tetraquarks $T_{QQ}$ via matching the model with the (measured or lattice)
mass data of the heavy hadrons. Using numerical diagonalization of
chromomagnetic interaction we predict that the ground-state masses of the
doubly charmed tetraquark $T_{cc}$ (in color rep. $\bar{3}_{c}\otimes 3_{c}$%
) with $IJ^{P}=01^{+}$ is $3879.2$ MeV, about $3$ MeV above the $D^{\ast
}D^{0}$ threshold, and that of the $T_{cc}^{\ast }$ (in $6_{c}\otimes \bar{6}%
_{c}$) with $IJ^{P}=01^{+}$ is about $4287.6$ MeV. The former prediction is
in consistent with the measured mass $3874.7\pm 0.05$ MeV of the doubly
charmed tetraquark $T_{cc}(1^{+})$ discovered by LHCb \cite{LHCb:2021vvq}.
The same method is applied to the systems $T_{bb}$ and $T_{bc}$ to compute
of the ground-state masses of the doubly bottom tetraquarks $T_{bb}$ and the
bottom-charmed tetraquarks $T_{bc}$ with $J^{P}=0^{+},1^{+},2^{+}$ and $%
I=0,1 $.

As the observed tetraquark $T_{cc}^{+}(3875)$ is very narrow and near
threshold, we suggest that the $J^{P}=1^{+}$ tetraquark $T_{cc}^{+}$($3875$)
observed by LHCb is the lowest state of the doubly-charmed tetraquark, being
a resonance state of the four-quark system $cc\bar{u}\bar{d}$ with color
configuration of the $\bar{3}_{c}\otimes 3_{c}$ dominately. Further, there
exists a new color partner $T_{cc}^{+}$($cc\bar{u}\bar{d}$, $6_{c}\otimes \bar{6}%
_{c}$) of the tetraquark $T_{cc}^{+}$($3875$) consisting of the $cc\bar{u}%
\bar{d}$ with dominate color configuration of $6_{c}\otimes \bar{6}_{c}$, having
the same $IJ^{P}=01^{+}$ but the mass about $4288$ MeV. A further
experimental search similar to the LHCb experiment \cite{LHCb:2021vvq} around
the inviariant-mass of the final states about $408$ MeV higher than the $%
T_{cc}^{+}$($3875$) can test our prediction.

The chromomagnectic-mixing enters generally more or less for the $%
IJ^{P}=01^{+}$ and $IJ^{P}=1(0^{+},1^{+})$ states of all DH tetraquarks $%
T_{QQ}$, for which the novel $6_{c}\otimes \bar{6}_{c}$ color state of the
tetraquarks occupies the higher mass levels. For the bottom-charmed
tetraquarks $T_{bc}$ with $IJ^{P}=01^{+}$, the chromomagnetical coupling are
strong enough among three color-spin states that some of them are mixed
substantially between the color configurations $\bar{3}_{c}\otimes 3_{c}$ and $%
6_{c}\otimes \bar{6}_{c}$.

We remark that the heavy quark pair in diquark $QQ$ in $\bar{3}_{c}$ may
(when they are very heavy) tend to stay close to each other to form a
compact core due to the strong Coulomb interaction, while it is also
possible (when $m_{Q}$ is comparable with $1/\Lambda _{QCD} \approx 3$ GeV$%
^{-1}$ ) that $Q$ attracts $\bar{q}$ to bind them into a colorless
clustering (in $1_{c}$) and other pair binds into another colorless
clustering (in $1_{c}$), resulting in a molecular system $(Q\bar{q}%
)(Q^{\prime }\bar{q}^{\prime })$. In the former case, DH tetraquarks mimic
the helium-like QCD-atom, while they resemble the hydrogen-like QCD
molecules in the later case. The possible molecule configurations of two
heavy mesons or mixture of molecules in the tetraquark systems are of
interest and remains to be explored.

\textbf{Acknowledgments }

D. J thanks Xiang Liu for discussions. D. J. is supported by the National
Natural Science Foundation of China under the no. 12165017.

\textrm{\ }

\end{document}